\documentclass[aps,prl,groupedaddress,nofootinbib,showpacs,twocolumn]{revtex4-1}

\usepackage{graphicx}
\usepackage[labelfont=small]{subcaption}
\usepackage{amsmath}

\graphicspath{{figures/}}

\newcommand{\dist}[1]{f_{#1}(#1)}

\newcommand{\matc}{\mathrm{ce}}
\newcommand{\mats}{\mathrm{se}}
\newcommand{\matcp}{\dot{\mathrm{ce}}}
\newcommand{\matsp}{\dot{\mathrm{se}}}
\begin{document}

\title{Superstatistical energy distributions of an ion in an ultracold buffer gas}
 
\author{I. Rouse}
\affiliation{Department of Chemistry, University of Basel, Basel, Switzerland}
\author{S. Willitsch}

\affiliation{Department of Chemistry, University of Basel, Basel, Switzerland}

\date{\today}

\begin{abstract}
An ion in a radiofrequency ion trap interacting with a buffer gas of ultracold neutral atoms is a driven dynamical system which has been found to develop a non-thermal energy distribution with a power law tail. The exact analytical form of this distribution is unknown, but has often been represented empirically by q-exponential (Tsallis) functions. Based on the concepts of superstatistics, we introduce a framework for the statistical mechanics of an ion trapped in an RF field subject to collisions with a buffer gas. We derive analytic ion secular energy distributions from first principles both neglecting and including the effects of the thermal energy of the buffer gas. For a buffer gas with a finite temperature, we prove that Tsallis statistics emerges from the combination of a constant heating term and multiplicative energy fluctuations. We show that the resulting distributions essentially depend on experimentally controllable parameters paving the way for an accurate control of the statistical properties of ion-atom hybrid systems.
\end{abstract}

\pacs{ }

\maketitle

%%%%% INTRODUCTION

The advent of hybrid systems of cold ions immersed in ultracold neutral atoms has opened up new perspectives for exploring two- and many-body effects in a regime intermediate between strong ion-ion and weak neutral-neutral couplings \cite{haerter14a, sias14a, willitsch15a}. A range of applications in atomic, molecular and chemical physics has recently emerged including studies of ion-neutral collisions and chemical reactions at very low energies \cite{zipkes10a, schmid10a, hall11a, rellergert11a, hall12a}, of many-body physics in dense systems \cite{harter12a, haerter13c} and of the quantum dynamics of an ion under the influence of an ultracold buffer gas \cite{ratschbacher13a, meir16a}. 

Ion-atom hybrid systems are realized by superimposing cold ions in a radiofrequency (RF) trap with trapped ultracold atoms \cite{haerter14a, sias14a, willitsch15a}. RF traps use rapidly oscillating electric fields to dynamically confine the ions. In an adiabatic regime \cite{gerlich92a}, the resulting motion of an ion can be represented as a thermal component (``secular motion'') superimposed by small-amplitude oscillations at the RF frequency (``micromotion'') \cite{gerlich92a}. In a hybrid trap, the ion undergoes frequent collisions with neutral atoms which disrupt its motion and lead to energy exchange between the secular motion and the RF field \cite{major68a, moriwaki92a, devoe09a,zipkes11a,chen14a, hoeltkemeier16a}. 

These processes lead to a distortion of the ion's secular-energy distribution from thermal (Boltzmann) to one better described by a power law at high energy \cite{levy96a, devoe09a, zipkes11a, chen14a}. The precise knowledge of the ion energetics is crucial for understanding the properties and dynamics of hybrid systems and their derived applications. Consequently, this problem has been the subject of intense recent research \cite{devoe09a, zipkes11a,chen14a,hoeltkemeier16a,hoeltkemeier16b}. In the high-energy limit, expressions for the mean energy and the power-law exponent have been derived \cite{chen14a}. The complete ion energy distribution has often been modeled \cite{devoe09a, biesheuvel16a, meir16a} by Tsallis (q-exponential) functions \cite{tsallis88a,tsallis09a}, 
\begin{equation} \label{eq:tsallis}
e_q(x) =(1 + (1-q_T) x)^{\frac{1}{1-q_T}}  
\end{equation}
for $q_T>1$ where $C_q$ is a normalization factor. $q_T$ is a parameter which characterizes the deviation from a standard exponential function which is recovered in the limit $q_T\rightarrow1$. However, the application of q-exponentials has remained empirical \cite{biesheuvel16a, hoeltkemeier16b, meir16b} since their first introduction for fitting numerical energy distributions \cite{devoe09a}. 

Using the formalism of superstatistics \cite{beck01a, beck03a}, we introduce a framework for the statistical mechanics of ion-atom hybrid systems. We derive analytic ion secular-energy distributions both neglecting and including the thermal energy of the ultracold buffer gas and confirm their validity by comparison with numerical simulations. For a buffer gas at zero Kelvin, we obtain an energy distribution with no steady-state and an exponential decay at high energies. For a buffer gas at finite temperature, we prove from first principles the emergence of Tsallis statistics thus vindicating its application in the present context. The energy distributions derived here depend on experimentally adjustable parameters which opens the door for a rational experimental control of the statistical properties of ion-atom hybrid systems. 

%%%%% Ion motion

The motion in each direction $r_j, j \in (x,y,z)$, of an ion in a quadrupole RF trap is given by the Mathieu differential equations,
\begin{equation}
\ddot r_j(\tau)  + [a_j - 2q_j \cos(2 \tau) ] r_j= 0 ,
\end{equation}
where $\tau = \Omega t/2$ and $q_j, a_j$ are the Mathieu stability parameters \cite{major05a}. 
In a hybrid system, the ion interacts with neutral atoms through a polarization potential. We treat the dynamics as a series of elastic collisions in the Langevin approximation with an energy-independent rate \cite{gioumousis58a,chen14a}. The velocity $\mathbf{v}$ of the ion after a collision is \cite{zipkes11a,chen14a},
\begin{equation} \label{eq:collisionAtomVel}
\mathbf{v'}  =  \frac{1}{1 + \tilde{m}}  \mathbf{v}   +  \frac{\tilde{m}}{1 + \tilde{m}}  \mathbf{v_n}  + \frac{\tilde{m}}{1 + \tilde{m}}   \mathrm{R} \cdot (\mathbf{v} -\mathbf{v_n} )
\end{equation}
where $\tilde{m} = m_n/m_i$ is the ratio of the atom's to the ion's mass, $\mathrm{R}$ is a rotation matrix \cite{arvo91a} and $\mathbf{v_n}$ is the velocity of the neutral atom.  Primes refer to post-collision quantities. As we have no control over the instantaneous velocities of the particles at the time of collisions, $\mathbf{v}$ and $\mathbf{v_n}$ are random variables.

From Eq.~\eqref{eq:collisionAtomVel}, the ion's secular energy after a collision can be derived to be (see supplemental material),
\begin{equation} \label{eq:energyPostCollision}
E' = \eta E + c_1 \sqrt{E \epsilon} + c_2 \epsilon,
\end{equation}
where $\epsilon = \frac{m_n}{2} |\mathbf{v_n}|^2$ is the kinetic energy of the neutral atom and $\eta,c_1,c_2$ are coefficients (see supplemental material). Assuming that the buffer gas density is uniform, these coefficients are independent of the values of $E$ and $\epsilon$. For an ion much hotter than the buffer gas $(E \gg \epsilon)$, we approximate $E' \approx \eta E$. The stability of the ion motion in the buffer gas with respect to runaway heating is determined by the distribution of $\eta$. As a rule, the motion is stable for a mass ratio $\tilde{m}\lesssim1.4$ for Mathieu parameters $q \ll 1$ \cite{moriwaki92a,devoe09a,zipkes11a,chen14a,hoeltkemeier16a}. 

\begin{figure}
    \centering
\includegraphics[width=.9\columnwidth]{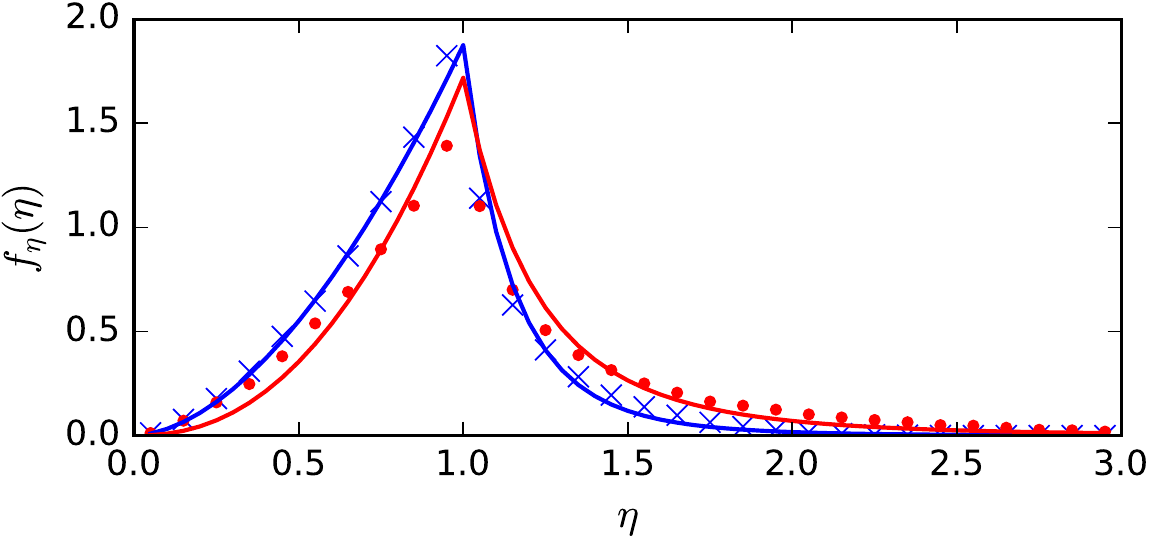}
    \caption{Distributions of the energy-transfer factor $\eta$ in ion-atom collisions for  $q = 0.1, \tilde{m} = 0.75$ (blue crosses) and $q = 0.5, \tilde{m} = 1.25$ (red points) starting from a thermal state with ion temperature $T_0=1$~mK. The points are binned normalized data from 100'000 numerical simulations of a collision. The lines represent an empirical asymmetric log-Lapace distribution, see text.
\label{fig:etaDist}}
\end{figure}

Fig.~\ref{fig:etaDist} shows numerical distributions $\dist{\eta}$ for the energy-transfer parameter $\eta$ for $q=0.1, \tilde{m}=0.75$ and $q=0.5, \tilde{m}=1.25$ which correspond to stable and unstable ion motions, respectively. The numerical simulations were performed following DeVoe's approach \cite{devoe09a}.
The solid lines in Fig.~\ref{fig:etaDist} correspond to log-Laplace distributions of the form \cite{kozubowski03a},
\begin{equation} \label{eq:empiricalEta}
\dist{\eta} =  \frac{1}{\delta} \frac{a_1 a_2}{a_1 + a_2}\begin{cases} 
      \left(\frac{\delta}{\eta} \right)^{a_1 + 1} & \eta \geq \delta \\
      \left(\frac{\eta}{\delta} \right)^{a_2 - 1} & 0 < \eta < \delta 
   \end{cases}
\end{equation}
with $a_1,a_2  > 0$ which have previously been used as to model processes involving multiplicative fluctuations \cite{kozubowski03a}. The parameter $\delta$ representing the maximum of the distribution was found to be $\approx 1$, reflecting the fact that most collisions result in little changes to the ion's energy. The values of $a_1$ and $a_2$ may be estimated by calculating the first and second moment $\langle\eta\rangle$ and $\langle\eta^2\rangle$, respectively, of the distribution using Eq. \eqref{eq:empiricalEta} and matching them to the expressions found numerically from Eq. \eqref{eq:energyPostCollision}.

Let us now assume that the ion is initially prepared in a thermal state at temperature $T_0$, as may be the situation after Doppler laser cooling \cite{metcalf99a,honerkamp02a}. 
The resulting distribution for the ion's initial energy $E_0$ is,
\begin{equation} \label{eq:energyInitialDistK}
\dist{E_0} = \frac{E_0^k \beta_0^{k+1} }{\Gamma(k+1)} e^{-E_0 \beta_0} ,
\end{equation}
where $\beta_0 = 1/(k_B T_0)$, $\Gamma$ is the Gamma function and the pre-exponential factor represents the density of states ($k = 2$ for a three-dimensional harmonic oscillator \cite{pethick01a}). 

We now consider the effects of collisions with the neutral atoms. Initially, we neglect their thermal energy and set $\epsilon = 0$ in Eq.~\eqref{eq:energyPostCollision} such that $E' = \eta E$. The resulting energy distribution can be written as \cite{riley10a},
\begin{align} \label{eq:energyMultiplication}
\dist{E'}&= \int_{\eta=0}^{\eta=\infty} \frac{1}{\eta} f_E(E'/\eta)  \dist{\eta} \mathrm{d} \eta \notag \\
&= \int_{\eta=0}^{\eta=\infty} \frac{1}{\eta}  \frac{(E'/\eta)^k \beta_0^{k+1} }{\Gamma(k+1)} e^{-(E'/\eta) \beta_0}   \dist{\eta} \mathrm{d} \eta  .
\end{align}
We first consider the case in which every collision multiplies the energy by a fixed amount, $\eta_c$. The distribution for $\eta$ is then given by a Dirac $\delta$ function,
\begin{equation}
\dist{\eta} = \delta(\eta - \eta_c) ,
\end{equation}
so that
\begin{equation} \label{eq:energyMultiplicationByConstant}
\dist{E'} = \frac{E'^k \beta_0^{k+1} }{\eta_c^{k+1}  \Gamma(k+1)} e^{-\frac{E' \beta_0}{\eta_c}}   .
\end{equation}
This is still a thermal distribution, except that it can now be written in terms of $\beta' = \beta_0/\eta_c$. 

We now generalise this approach to an arbitrary $\dist{\eta}$ by making the change of variables $\beta' = \beta_0 / \eta$ in Eq.~\eqref{eq:energyMultiplication},
\begin{equation} \label{eq:betaDistRelation}
\dist{E'}   = \int_{\beta'=0}^{\beta'=\infty} \frac{E'^k \beta'^{k+1} }{\Gamma(k+1)} e^{-E' \beta'}    \frac{\beta_0}{\beta'^2}f_\eta\left(\frac{\beta_0}{\beta'}\right)\mathrm{d} \beta' .
\end{equation}
The energy distribution after a collision can thus be represented by a superposition of thermal states. This problem can be treated within the formalism of superstatistics, i.e., the superpositions of several statistics as in our case the ones of $\eta$ and $E$ in Eq. (\ref{eq:energyMultiplication}) \cite{beck01a, beck03a, rouse15a}.

We can now define a distribution for $\beta'$,
\begin{equation} \label{eq:betaDist1}
\dist{\beta'}  = \frac{\beta_0}{\beta'^2}f_\eta\left(\frac{\beta_0}{\beta'}\right),
\end{equation}
which is used to recast Eq. (\ref{eq:betaDistRelation}) into the form
\begin{equation} \label{eq:energyAveragedOverBeta}
\dist{E'}   = \int_{\beta'=0}^{\beta'=\infty} \frac{E'^k \beta'^{k+1} }{\Gamma(k+1)} e^{-E' \beta'}   \dist{\beta'}  \mathrm{d} \beta'  .
\end{equation}
Eq. (\ref{eq:energyAveragedOverBeta}) has the form of a Laplace transform $\mathcal{L}$.
For general distributions $\dist{\beta}, \dist{\eta}$ one gets
\begin{equation} \label{eq:betaUpdate}
\dist{\beta'} = \int_{\eta=0}^{\eta=\infty} \eta f_\beta(\eta \beta') f_\eta(\eta) d\eta .
\end{equation}
Repeated application of Eq.~\eqref{eq:betaUpdate} and substitution into Eq.~\eqref{eq:energyAveragedOverBeta} can then be performed to obtain the energy distribution of an ion after $n$ collisions. 

Thus, we formulate a recurrence relation for $\beta$ after collision number $i$,
\begin{equation}
\beta_i = \beta_{i-1} / \eta_i.
\end{equation}
Since the ion is initially in a thermal state, we take $\beta_0$ to be constant. After $n$ collisions starting from $\beta_0$, we get 
\begin{equation}
\beta_n = \beta_0  \prod_{i=1}^{n} 1/ \eta_i .
\end{equation}
Each value of $\eta$ is assumed to be independently and identically distributed, and so by applying the central limit theorem the product $\prod_{i=1}^{n} 1/ \eta_i$ is log-normally distributed for large $n$ \cite{riley10a}. Hence, from Eq.~\eqref{eq:betaDist1} we write,
\begin{equation} \label{eq:betaDistNCollisionsLN}
\dist{\beta_n} =  \frac{1}{\sqrt{2 \pi n}  \sigma \beta_n} \exp[   -\frac{ (\ln \beta_n - \ln \beta_0 + n\mu)^2  }{2 n \sigma^2}  ]  ,
\end{equation}
where $\mu =    \langle  \ln \eta \rangle$ and $\sigma^2 =  \langle( \ln \eta )^2 \rangle - \langle \ln \eta \rangle^2$.

We now return to the energy distribution. By inserting Eq.~\eqref{eq:betaDistNCollisionsLN} into Eq.~\eqref{eq:energyAveragedOverBeta}, we obtain,
\begin{align} \label{eq:energyDistribution}
\dist{E_n} &=  \int_{\beta_n=0}^{\beta_n=\infty}  \frac{E_n^k \beta_n^{k+1} }{\Gamma(k)} e^{-E_n \beta_n} \notag \\
& \times \frac{1}{\sqrt{2 \pi n}  \sigma \beta_n} \exp[   -\frac{ (\ln \beta_n - \ln \beta_0 + n\mu)^2  }{2 n \sigma^2}  ]   \mathrm{d}\beta_n.
\end{align}
We use the Laplace integration method \cite{asmussen16a} to find an approximate analytical solution for $k=2$. We obtain
\begin{align} \label{eq:eDistNApprox}
&\dist{E_n} = \frac{\hat{\beta}^3 E_n^2 }{4 \sqrt{\hat{\beta} E_n n \sigma^2+1}}  \exp \left( -\hat{\beta} E_n \right)  \notag \\
& \times \left(\text{erf}  \left(\sqrt{\frac{\hat{\beta}E_n n \sigma^2+1}{2 n \sigma^2}}\right)  +1\right) \exp \left(  - \frac{n \sigma^2}{2} \left( \hat{\beta}E_n -2\right)^2  \right) ,
\end{align}
where $\hat{\beta}$ is the point at which the integrand of Eq. (\ref{eq:energyDistribution}) is maximal. In the high-energy limit for $k=0$, Eq. (\ref{eq:eDistNApprox}) has been shown to exhibit an exponential decay \cite{touchette05a, rabassa14a}. From the general property of the Laplace transform,
\begin{equation}
\mathcal{L}[ \beta^{k+1} \dist{\beta} ] = (-1)^{k+1} \frac{d^{k+1}}{d E^{k+1}} \mathcal{L}[ \dist{\beta} ] ,
\end{equation}
follows that if the high-energy behavior for $k=0$ is an exponential decay, then this holds true for any integer value of $k$. Thus, we conclude that a purely multiplicative model of the heating process does not lead to Tsallis statistics which is characterized by a power-law tail for the distribution at high energies.

\begin{figure}[b]
    \centering
\includegraphics[width=.9\columnwidth]{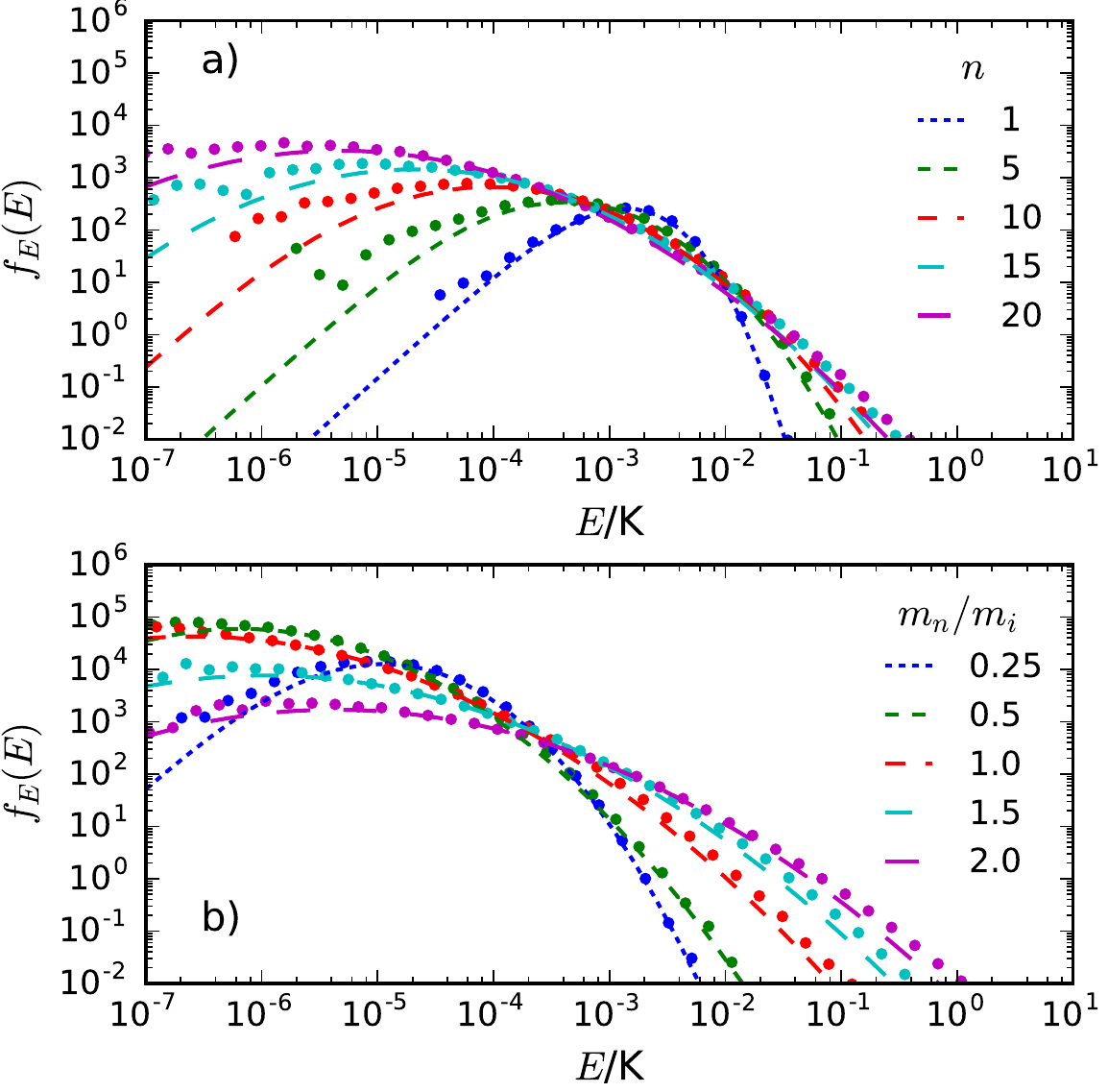}
    \caption{(a) Energy distributions of an ion in a RF trap after $n$ collisions with a neutral buffer gas at zero Kelvin with a mass ratio $\tilde{m}=m_n/m_i=1.5$. \\ (b) The ion energy distribution after 25 collisions at a range of mass ratios. The lines show corresponding energy distributions computed with Eq.~\eqref{eq:eDistNApprox}. The points show numerical data sampled after 100'000 simulations. \label{fig:collisionSeries}}
\end{figure}

In order to test the validity of Eq.~\eqref{eq:eDistNApprox}, a series of simulations were performed at a buffer gas temperature $T=0$~K and varying the mass ratio or number of collisions. The results are plotted in Fig.~\ref{fig:collisionSeries} along with the distributions computed from Eq.~\eqref{eq:eDistNApprox}. The $\mu$ and $\sigma$ parameters were computed from numerical distributions $\dist{\eta}$ such as the ones shown in Fig. \ref{fig:etaDist}. At low collision numbers, the agreement is generally poor, which is expected due to the assumption in the derivation of Eq. (\ref{eq:eDistNApprox}) that the central limit theorem can be applied. Moreover, for all collision numbers, the agreement is less good at low energies due to the Laplace integration method being valid only in the limit $E\rightarrow\infty$. However, for higher energies and numbers of collisions, Eq.~\eqref{eq:eDistNApprox} becomes an increasingly better representation of the simulated data. 

\begin{figure}
\centering
\includegraphics[width=.9\columnwidth]{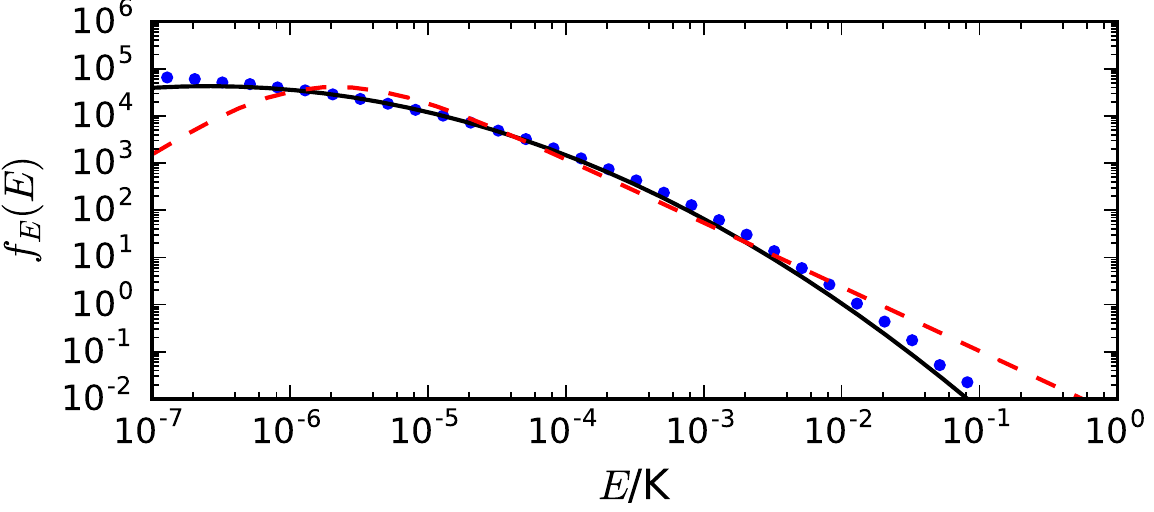}
\caption{Comparison between the ion energy distribution Eq. (\ref{eq:eDistNApprox}) for a buffer gas at 0~K (black dashed line) and a Tsallis distribution (red dashed line) for an ion in a Paul trap after 25 collisions with a mass ratio 1.0. The points represent numerical data sampled from 100'000 simulations.}
\label{fig:tsallisAsymptoticCompare}
\end{figure}

For comparison, the numerical data for 25 collisions at a mass ratio of 1.0 is presented in Fig.~\ref{fig:tsallisAsymptoticCompare} together with the distribution predicted using Eq.~\eqref{eq:eDistNApprox}. The red dashed line represents a Tsallis distribution obtained from a maximum-likelihood estimation (MLE) to the numerical data. It can be clearly seen that Tsallis statistics is a poor match for a buffer gas at zero Kelvin, while Eq.~\eqref{eq:eDistNApprox} provides much better agreement. 

Neither the energy nor the $\beta$ distributions, Eq. (\ref{eq:eDistNApprox}) and Eq. (\ref{eq:betaDistNCollisionsLN}), respectively, converge to a steady state with increasing $n$. This is a known property of an unbounded multiplicative random walk, and in the present case results from the neglect of the temperature of the buffer gas allowing the ion to reach an arbitrarily low temperature \cite{levy96a, sornette97a}.

For a buffer gas at a finite temperature, we have to adopt a different procedure as the change of the ion energy following a collision is no longer a purely multiplicative process, see Eq. \eqref{eq:energyPostCollision}. Because the buffer gas velocity distribution is isotropic, the $c_1$ coefficient averages to zero such that it can be neglected. We are thus left with $E'=\eta E+c_2\epsilon$.
Assuming again that the ion's energy distribution can be represented as a superposition of thermal states as in Eq. (\ref{eq:betaDistRelation}), it follows that $\langle E\rangle=(1+k)k_B\langle T \rangle$. This suggests that we can rephrase the problem of finding an energy distribution to one of finding the underlying temperature distribution.  We approximate that the contributions from $\epsilon$ in Eq. (\ref{eq:energyPostCollision}) can be treated as a constant source of heating proportional to the temperature of the buffer gas $T_a$ which ensures that the ion's steady-state temperature is non-zero. This is a good approximation if the thermal fluctuations of the buffer gas are much smaller than the ones of the ion. The ion temperature after a collision is then,
\begin{equation} \label{eq:recurrenceRelationWithKappa}
T_{i}= \eta_i T_{i-1} + \kappa T_a ,
\end{equation}
where $\kappa$ is a heating coefficient (see supplemental material). 

To find the required temperature distribution, we solve the recurrence relation Eq. (\ref{eq:recurrenceRelationWithKappa}). The mathematical solution of this problem has been outlined in Refs. \cite{sornette97a, biro05a} and leads to a gamma distribution for $\beta$:
\begin{equation} \label{eq:betaDistClean}
\dist{\beta} = \frac{1}{\beta \Gamma (n_T)}   e^{-\frac{\beta n_T}{\langle \beta  \rangle}} \left(\frac{\beta n_T}{\langle \beta  \rangle}\right)^{n_T} .
\end{equation}
Multiplying by the density of states and applying the Laplace transform we obtain the ion energy distribution,
\begin{equation} \label{eq:tsallisEnergyDist}
f_{E,T}(E) = \left(\frac{n_T}{\langle\beta\rangle}\right)^{-k-1}  \frac{\Gamma (k+n_T+1)}{\Gamma (k+1) \Gamma (n_T)}     \frac{E^k}{ \left(\frac{\langle\beta\rangle E}{n_T}+1\right)^{k + n_T + 1}}.
\end{equation} 
The parameter $n_T$ can be obtained from the condition \cite{sornette97a}
\begin{equation} \label{eq:nuDef}
\int_{\eta=0}^{\eta=\infty} f_\eta(\eta) \eta^{n_T} d \eta = 1.
\end{equation}
This integral may be solved numerically, or alternatively we make use of the empirical distribution Eq. (\ref{eq:empiricalEta}). From substituting Eq.~\eqref{eq:empiricalEta} into Eq.~\eqref{eq:nuDef}, we obtain,
\begin{equation} \label{eq:nuApproxPowerLawDist}
n_T = a_1 - a_2 = \frac{\langle \eta \rangle - 4 \langle \eta^2 \rangle + 3 \langle \eta \rangle \langle \eta^2 \rangle}{\langle \eta \rangle - 2 \langle \eta^2 \rangle + \langle \eta \rangle \langle \eta^2 \rangle} .
\end{equation}
assuming $\delta = 1$ in Eq.~\eqref{eq:empiricalEta}.
To fully characterize Eq. (\ref{eq:tsallisEnergyDist}), we also require the value for $\langle \beta \rangle$. From Eq. (\ref{eq:recurrenceRelationWithKappa}), it follows that
\begin{equation} \label{eq:meanTFromRecurrence}
\langle T \rangle = \langle \eta \rangle \langle T \rangle + \kappa T_a = \frac{\kappa T_a}{1 - \langle \eta \rangle} ,
\end{equation}
Averaging $T = 1/(k_B \beta)$ over Eq.~\eqref{eq:betaDistClean}, we get,
\begin{equation} \label{eq:meanTFromGamma}
\langle T \rangle = \frac{1}{k_B \langle \beta \rangle} \frac{n_T}{n_T-1} .
\end{equation}
Equating Eqs.~\eqref{eq:meanTFromRecurrence} and \eqref{eq:meanTFromGamma} we find,
\begin{equation} \label{eq:BetaMean}
\langle \beta \rangle = \frac{1}{k_B \kappa T_a } \frac{n_T}{n_T -1 } (1 - \langle \eta \rangle )  .
\end{equation}
This derivation is only valid for $n_T > 1$ and $\langle \eta \rangle < 1$. If either of these conditions is not met, the mean temperature diverges because the ion motion becomes unstable.

The distribution Eq.~\eqref{eq:tsallisEnergyDist} has the form of a q-exponential Eq. (\ref{eq:tsallis}) multiplied by a $E^k$ term. For  $k=0$ (one-dimensional), it reduces to the standard q-exponential, and for $k=2$ (three-dimensional) it is equivalent to the form used in Ref.~\cite{meir16b}, if we set their exponent $n = n_T + 3$. We have therefore shown that Tsallis statistics are physically meaningful for the present problem under the condition that the variance of the thermal fluctuations are sufficiently small so that the additive noise due to the thermal energy of the atoms can be approximated as a constant.
\begin{figure}
\centering
\includegraphics[width=.9\columnwidth]{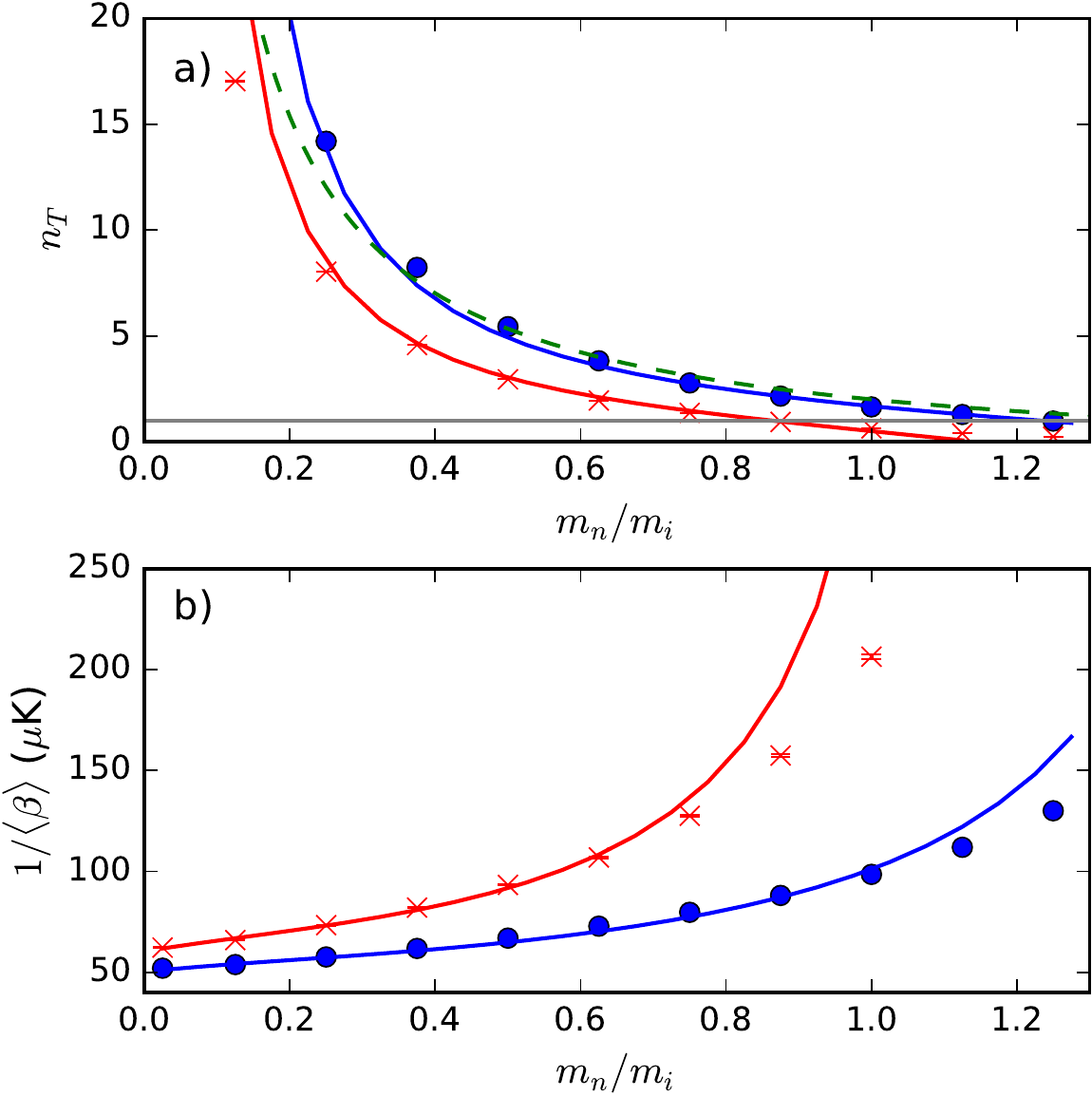}
    \caption{(a) Tsallis parameter $n_T$ at Mathieu parameter $q= 0.1$ (blue circles) and $q=0.5$ (red crosses) calculated by a maximum likelihood estimation (MLE) of a Tsallis function to the steady-state ion-energy distribution obtained from numerical simulations (100'000 trials per point). The blue and red lines show the predictions using Eq.~\eqref{eq:nuApproxPowerLawDist}. The green dotted line indicates the approximate result for $q < 0.4$ from Ref.~\cite{chen14a} and the grey horizontal line indicates the critical exponent  $n_T=1$ below which the mean energy is undefined. (b) As (a) for $1/\langle \beta \rangle$. Error bars correspond to the standard errors of the MLE values and are plotted when larger than the size of the symbols.
\label{fig:parameterComparison}}
\end{figure}

Fig.~\ref{fig:parameterComparison} shows a comparison of the MLE values of the parameters $n_T$ and $1/\langle\beta\rangle$ extracted from numerical simulations with their predictions from Eqs. (\ref{eq:nuApproxPowerLawDist}) and (\ref{eq:BetaMean}), respectively. Below the critical mass ratio given by the intersection of the curves with the grey horizontal line in Fig. \ref{fig:parameterComparison} (a), the ion motion is stable. Up to near this point, the predictions for both parameters are very close to the values extracted from numerical data, vindicating the assumptions leading to the derivation of Eq. (\ref{eq:betaDistClean}). Above the critical mass ratio, the predicted mean $\langle \beta \rangle$ becomes increasingly inaccurate as a result of energy correlations between different coordinate axes not accounted for in the present model (see \cite{chen14a} and supplemental material). 

From Eq. \eqref{eq:tsallisEnergyDist} and \eqref{eq:nuDef}, it becomes clear that the energy distribution and therefore the statistical properties of the ion depend on the buffer gas temperature and the distribution $\dist{\eta}$. The latter depends on system parameters such as the atom-ion mass ratio and the Mathieu parameters of the trap which are defined in advance by the experimenter. By varying these parameters, $\dist{\eta}$ and therefore the Tsallis distribution Eq. \eqref{eq:tsallisEnergyDist} can be tuned in a deterministic manner allowing for a control of the statistical properties of the system.

%Based on the concepts of superstatistics, we have found analytical expressions for the secular energy distributions of an ion in a RF trap interacting with a neutral buffer gas of both zero and finite temperatures. We have proven from first principles that Tsallis statistics is a physically meaningful model for ion-atom hybrid systems which arises from multiplicative fluctuations of the ion energy due to collisions bounded by the finite temperature of the buffer gas. We have derived analytical expressions for the resulting energy distributions obtained and showed that they essentially depend on experimentally controllable parameters. Thus, the current results open the door to accurately adjust and control the statistical properties of an ion in a hybrid system.

Beyond the current application, the formalism developed here represents a general framework for describing the statistical mechanics of an ion in a buffer gas which can be used to, e.g., compute thermodynamic functions \cite{beck11a}. The present treatment can also be extended to localized buffer gases. These developments will be reported elsewhere.

\begin{acknowledgments}
We acknowledge funding from the Swiss Nanoscience Institute project P1214 and the Swiss National Science Foundation as part of the National Centre of Competence in Research, Quantum Science \& Technology (NCCR-QSIT) and grant nr. 200021\_156182.
\end{acknowledgments}

%\bibliography{buffergas_cooling_extra,Main-Aug2016}
%

\appendix
\section{Supplemental Material}
\subsection{Ion motion} \label{sm:ionMotion}

The treatment of the motion of an ion in a quadrupole RF trap is given in standard texts, e.g., Ref.  \cite{major05a}. For ready reference, we repeat here the salient points. The equation of motion in each direction $r_j, j \in (x,y,z)$, of an ion in a quadrupole RF trap is given by the Mathieu differential equation ,
\begin{equation}
\ddot r_j(\tau)  + [a_j - 2q_j \cos(2 \tau) ] r_j= 0 ,
\end{equation}
where $\tau = \Omega t/2$ and $q_j, a_j$ are the Mathieu parameters for axis $j$. For stable ion trajectories, the solution can be written as a sum of even (ce) and odd (se) Mathieu functions,
\begin{equation} \label{eq:ionMotion2Mathieu}
r_j(\tau) = A_j \cos(\phi_{j0}) \mathrm{ce}(a_j,q_j,\tau) - A_j \sin(\phi_{j0}) \mathrm{se}(a_j,q_j,\tau) ,
\end{equation}
with amplitude $A_j$ and initial phase $\phi_{j0}$. The velocity of the ion is obtained to be 
\begin{equation} \label{eq:ionVelocity2Mathieu}
\dot{r}_j(\tau) = A_j \cos(\phi_{j0}) \mathrm{\dot{ce}}(a_j,q_j,\tau) - A_j \sin(\phi_{j0}) \mathrm{\dot{se}}(a_j,q_j,\tau),
\end{equation}
with,
\begin{equation}
\dot{\mathrm{ce}}(a_j,q_j,\tau) = - \sum_{m=-\infty}^{m=\infty} c_{2m,j} (\beta_j + 2m) \sin( (\beta_j + 2m) \tau  ) ,
\end{equation}
and the equivalent result for $\dot{\mathrm{se}}$. Here, the $c_{2m,j}$ are the coefficients of a Fourier expansion of ce and $\beta_j \approx \sqrt{(a_j + q_j^2/2)}$ for small $q_j$\cite{major05a}. By considering only the lowest order term $m=0$, we obtain the secular position of the ion as a function of $t$,
\begin{equation} \label{eq:ionMotion1Secular}
\tilde{r_j}(t) = A_j c_{0,j} \cos[ \beta_j \Omega / 2 t  + \phi_{j0}] ,
\end{equation}
which represents a harmonic oscillation of amplitude $\tilde{A}_j = A_j c_0$, frequency $\omega_j = \beta_j \Omega /2$ and initial phase $\phi_{j0}$. The secular velocity can be found through differentiation of the secular position, and from these we can define the secular energy,
\begin{equation} \label{eq:secularEnergyOneAxis}
\tilde{E}_j  =  \frac{m_i}{2}  \omega_j^2 \tilde{A}_j^2 = \frac{m_i}{2} \frac{\Omega^2}{4} \beta^2_j c_{0j}^2 A_j^2 .
\end{equation}
We define the total secular energy of the ion to be,
\begin{equation}
E = \tilde{E}_x + \tilde{E}_y + \tilde{E}_z = \frac{m_i}{2}  (\omega_x^2 \tilde{A}_x^2+\omega_y^2 \tilde{A}_y^2+\omega_z^2 \tilde{A}_z^2)   ,
\end{equation} 
which is the energy of a 3D harmonic oscillator with frequencies $\omega_{x,y,z}$.

\section{Derivation of post-collision ion energy} \label{sm:etaFunction}
Here, we outline a proof that the ion's energy after a micromotion-interrupting collision is given by Eq.~(4) in the main text, i.e., $E' = \eta E + c_1 \sqrt{E \epsilon} + c_2 \epsilon$, and a method to derive the coefficients $\eta, c_1, c_2$ in this expression. We shall also discuss how the random variables involved may be generated so that these coefficients may be sampled for comparisons with numerical simulations.

The velocity after a collision is given by Eq.~(3) of the main text, i.e.,
\begin{equation} \label{eq:collisionAtomVel2}
\mathbf{v'}  =  \frac{1}{1 + \tilde{m}}  \mathbf{v}   +  \frac{\tilde{m}}{1 + \tilde{m}}  \mathbf{v_n}  + \frac{\tilde{m}}{1 + \tilde{m}}   \mathrm{R} \cdot (\mathbf{v} -\mathbf{v_n} ) ,
\end{equation}
In terms of the amplitudes $A_j$ and the initial secular phases $\phi_{j0}$, the components of the velocity are given by,
\begin{equation} \label{eq:velocitySeparatePhases}
v_j(t) =  A_j  (\cos(\phi_{j0}) \dot{\matc}(a_j,q_j,\tau) -\sin(\phi_{j0}) \dot{\mats}(a_j,q_j,\tau)  ) ,
\end{equation}
with the equivalent for the post-collision velocity, where we use $A_j', \phi_{j0}'$ to indicate the post-collision values.  To proceed, we eliminate $\phi_{j0}'$. We approximate that the ion does not move during the collision and so by equating the positions before and after we can find an expression for $\phi_{j0}'$. Thus, we take, 
\begin{equation} \label{eq:positionSeparatePhases}
r_j(t) =  A_j  (\cos(\phi_{j0}) \matc(a_j,q_j,\tau) -\sin(\phi_{j0}) \mats(a_j,q_j,\tau)  ) ,
\end{equation}
then expand the Mathieu functions in a Fourier series, and apply standard trigonometric addition formulae to produce,
\begin{equation} \label{eq:positionCombinedPhase}
r_j(t) =  A_j  \sum_{m=-\infty}^{m=\infty} c_{2m,j} \cos[\phi_{j0}  +(\beta + 2 m) \tau ] .
\end{equation}
Next, we apply the harmonic addition theorem to rewrite this as a single trigonometric function \cite{oo12a},
\begin{equation} \label{eq:positionCombined}
r_j(t) =  A_j  C_j \cos[\phi_{j0}  + \delta_{\tau,j} ] ,
\end{equation}
where,
\begin{equation}
C_j^2 = \sum_{m=-\infty}^{m=\infty} \sum_{n=-\infty}^{n=\infty}  c_{2m,j}  c_{2n,j} \cos( 2(m-n) \tau),
\end{equation}
and
\begin{equation}
\tan \delta_{\tau,j} = \frac{\sum_{m=-\infty}^{m=\infty} c_{2m,j} \sin[(\beta + 2m)\tau]}{\sum_{m=-\infty}^{m=\infty} c_{2m,j} \cos[(\beta + 2m)\tau]} = \frac{\mats(a_j,q_j,\tau)}{\matc(a_j,q_j,\tau)} .
\end{equation}
By equating the position before and after the collision we find,
\begin{equation}
A_j'  C_j \cos[\phi_{j0}'  + \delta_{\tau,j} ] = A_j  C_j \cos[\phi_{j0}  + \delta_{\tau,j} ] .
\end{equation}
Dividing by $C_j$ and rearranging for the post-collision secular phase gives,
\begin{equation}
\phi_{j0}'   = \cos^{-1} \left[ \frac{ A_j}{A_j'}  \cos\left(\phi_{j0}  + \delta_{\tau,j} \right) \right] - \delta_{\tau,j} .
\end{equation}
We may now substitute this into Eq.~\eqref{eq:velocitySeparatePhases} to find the post-collision velocities as a function of $A_j', A_j, \phi_{j0}$, 
\begin{align}
v_j' = &\frac{\Omega}{2}  [\matcp(a_j,q_j,\tau) \left(A_j \cos \delta_{\tau,j} \cos (\delta_{\tau,j}+\phi_{j0}) + s_j \sin \delta_{\tau,j}\right) \notag \\ &+\matsp(a_j,q_j,\tau) \left(A_j \sin \delta_{\tau,j}  \cos (\delta_{\tau,j}+\phi_{j0})- s_j \cos \delta_{\tau,j} \right) ] ,
\end{align}
where
\begin{equation}
s_j = \sqrt{A_j'^2-A_j^2 \cos ^2(\delta_{\tau,j}+\phi_{j,0})} .
\end{equation}
This result is substituted into Eq.~\eqref{eq:collisionAtomVel2} for the components of $\mathbf{v'}$, and likewise Eq.~\eqref{eq:velocitySeparatePhases} is substituted in for the components of $\mathbf{v}$. Thus we obtain a vector equation linking the post-collision amplitudes of motion to the pre-collision amplitude and secular phases, the neutral velocity components, and the random rotation matrix $\mathrm{R}$ in Eq. \eqref{eq:collisionAtomVel2}. The equation can then be simplified such that each component of the left-hand side is given by $s_j$ and so we write,
\begin{equation} \label{eq:simpleVecEquation} 
 \mathbf{s} = \mathbf{v_R} ,
\end{equation}
where $\mathbf{s}^T = (s_x,s_y,s_z)$ and $\mathbf{v_R}$ contains the remaining terms, none of which depend on $A_j'$. 
Next, we must convert this to a set of equations for the secular energy.  These energies are related to the amplitudes by Eq.~\eqref{eq:secularEnergyOneAxis}; by inspection we will need to square the equations in order to extract the secular energy, since they are presently linear in the amplitude. We take the outer product of $\mathbf{s}$ with itself  resulting in,
\begin{equation} \label{eq:postCollisionAmplitudeMatrix}
\mathbf{s}\otimes\mathbf{s} = \begin{pmatrix}
 s_x^2&   s_x s_y &   s_x s_z \\
s_x s_y &    s_y^2 &  s_y s_z \\
   s_x s_z&   s_y s_z &    s_z^2
\end{pmatrix} .
\end{equation}
 The diagonal terms of Eq.~\eqref{eq:postCollisionAmplitudeMatrix} are given by,
\begin{equation}
A_j'^2-A_j^2 \cos ^2(\delta_{\tau,j}+\phi_{j,0}) \propto \tilde{E_j'} - \tilde{E_j}\cos ^2(\delta_{\tau,j}+\phi_{j,0}), 
\end{equation}
where the proportionality factor can be found from Eq.~\eqref{eq:secularEnergyOneAxis}. By equating these terms to the corresponding term in the matrix given by $\mathbf{v_R} \otimes \mathbf{v_R}$ we obtain an expression for $\tilde{E_j'}$ in terms of the amplitudes and secular phases for each axis before the collision, the elements of the rotation matrix, $\tau$ and the velocity of the neutral atom.  We may use these to find the steady-state energy by averaging over the collision parameters (e.g. $x_1, \phi_j, \tau$) and requiring that in the steady state $\langle \tilde{E_j'} \rangle = \langle \tilde{E_j} \rangle$ to generate three simultaneous equations of the general form,
\begin{equation}
\langle \tilde{E_j} \rangle = \eta_{jx} \langle \tilde{E_x} \rangle + \eta_{jy} \langle \tilde{E_y} \rangle + \eta_{jz} \langle \tilde{E_z} \rangle +  \alpha E_n,
\end{equation}
where $\alpha E_n$ is the fraction of the neutral energy transferred to the ion, and the $\eta_{ij}$ are coefficients describing the combination of the transfer of energy between the axes and the random fluctuation of the energy due to micromotion interruption. The set of three equations then may then be solved to find the three mean steady-state energy components $\langle \tilde{E_j} \rangle$. This approach leads to essentially the same set of equations as is found in Ref.~\cite{chen14a}, but in terms of the secular rather than the time-averaged energies. Finally, we note that the off-diagonal elements may be averaged over in the same manner to determine quantities such as $\langle A_x A_z \rangle$ which contain information about the correlations between the motion along each axis and may be of use in future investigations. 

Instead of directly averaging over these quantities to obtain the mean energies, we may also use them to prove that the form of Eq.~(4) in the main text is correct, and to extract the multiplicative coefficient $\eta$. Returning to the set of three equations defined by the diagonal elements of the matrices, we convert the system to a form of spherical coordinates defined by,
\begin{equation}
\begin{matrix} 
A_x c_{0,x} \beta_x \Omega/2 = \rho \cos(\phi_{\rho}) \sin(\theta_{\rho}) \\
A_y c_{0,y} \beta_y \Omega/2= \rho \sin(\phi_{\rho}) \sin(\theta_{\rho}) \\
A_z c_{0,z} \beta_z \Omega/2= \rho  \cos(\theta_{\rho}) ,
\end{matrix}
\end{equation}
with the the two angles defined in the interval $[0,\pi/2)$, and an equivalent transformation applied to the primed quantities. The advantage of this coordinate system is that it simplifies the factoring of the total energy given by $E = \frac{m_i}{2} \rho^2$ from the expressions, since each secular energy component $\tilde{E_j}$ is proportional to $\rho^2$ and a function of $\phi_\rho,\theta_\rho$. For $v_n$, standard spherical coordinates may be used,
\begin{equation}
\begin{matrix} 
v_{nx} = |v_n| \cos(\phi_{n}) \sin(\theta_{n}) \\
v_{ny} = |v_n| \sin(\phi_{n}) \sin(\theta_{n}) \\
v_{nz} = |v_n| \cos(\theta_{n}) ,
\end{matrix}
\end{equation}
with $ \epsilon = \frac{m_n}{2}|v_n|^2$.  Performing the conversion to spherical coordinates is then followed by rearranging each equation such that the terms proportional to $\rho'$ are on the left hand side, and summing the three equations together. This produces an equation of the form,
\begin{equation}
\rho'^2 = \eta \rho^2 + \tilde{c_1} \rho |v_n| + \tilde{c_2} |v_n|^2 .
\end{equation}
In terms of the energies this is, 
\begin{equation} \label{eq:appendixEnergyRelation}
E' = \eta  E + c_1 \sqrt{E \epsilon} + c_2  \epsilon ,
\end{equation}
where the coefficients have been redefined to include the factors of $m_i$ and $m_n$ to match the result given in Eq.~(4) of the main text. 

The factor $\eta$ is a function of nine random variables -- the three initial secular phases $\phi_{0,j}$, the time $\tau$, the three variables $x_1,x_2,x_3$ used in the random rotation matrix given in Ref.~\cite{arvo91a}, and $\theta_\rho,\phi_\rho$ which describe the relative distribution of the secular energy between the three axes. Of these, the $\phi_{0,j}$ are uniformly distributed on $[0,2\pi)$  for a homogenous buffer gas, and the three rotation matrix variables $x_1,x_2,x_3$ are uniformly distributed on $[0,1)$.  The angles $\theta_\rho$ and $\phi_\rho$ are given by, 
\begin{equation}
\phi_\rho = \tan^{-1} \left( \frac{A_y c_{0,y} \beta_y}{A_x c_{0,x} \beta_x} \right) = \tan^{-1} \left( \frac{ \tilde{A}_y \omega_y}{ \tilde{A}_x \omega_x} \right) ,
\end{equation}
and,
\begin{equation} \label{eq:thetaRhoDef}
\theta_\rho = \cos^{-1} \frac{A_z c_{0,z} \beta_z}{\sqrt{ (A_x c_{0,x} \beta_x)^2 + (A_y c_{0,y} \beta_y)^2 + (A_z c_{0,z} \beta_z) ^2  }} .
\end{equation}

In the ideal case, the temperature is a constant and is equal for each axis. Under these conditions, the probability distribution may be found analytically by starting from the thermal distribution for $A_j$ (see Eq.~\eqref{eq:thermalAmplitudeDist} below) and applying the standard methods for finding functions of random variables \cite{riley10a}. The final results are,
\begin{equation}
\dist{\phi_\rho} = \sin(2\phi_\rho), 0 < \phi_\rho < \pi/2
\end{equation}
and
\begin{equation}
\dist{\theta_\rho} = 4 \cos \theta_\rho \sin^3 \theta_\rho , 0 < \theta_\rho < \pi/2 .
\end{equation}
Random sampling of these distributions may be achieved through the inverse transform method by taking a random variable $u_i$ uniformly distributed in $[0,1)$ :
\begin{equation}
\phi_\rho = \sin^{-1}( \sqrt{u_1}), 
\end{equation}
and
\begin{equation}
\theta_\rho = \sin^{-1}( u_2^{1/4}) .
\end{equation}
In practice, especially at higher mass ratios and values of the Mathieu $q$ parameter, this assumption breaks down, and the temperature for each axis is defined by a separate distribution. For linear RF traps with radial symmetry \cite{major05a}, the distributions for the radial $x$ and $y$ axes are identical and $\dist{\phi_\rho}$ is approximately unchanged, but $\dist{\theta_\rho}$ must be corrected to take into account the breakdown of equipartition between the $(x,y)$ and $z$ axes \cite{chen14a}. We will take each axis to have a different inverse temperature $\tilde{\beta_j}$, leading to the expected different mean energy for each axis. We approximate that, for a linear trap, $\tilde{\beta_x} = \tilde{\beta_y}$   and define $\xi = \tilde{\beta_x}/\tilde{\beta_z} = T_z / T_x$.  Under these conditions,  $\dist{\theta_\rho}$ can be re-derived as , 
\begin{equation} \label{eq:thetaRhoCorrected}
\dist{\theta_\rho} = \frac{4 \xi^2 \sin ^3(\theta_\rho) \cos (\theta_\rho)}{\left((1-\xi) \cos ^2(\theta_\rho)+\xi\right)^3} .
\end{equation}
For low mass ratios, the breakdown from equipartition is small and it suffices to set $\xi = \langle \xi \rangle \approx \tilde{E_z}/\tilde{E_x}$. At higher mass ratios, the increased correlation between the energy along each axis is such that $\langle \xi \rangle$ decreases less rapidly than predicted, and the higher-order moments of $\xi$ must be taken into account.  

\begin{figure}
    \centering
\includegraphics[width=0.9\columnwidth]{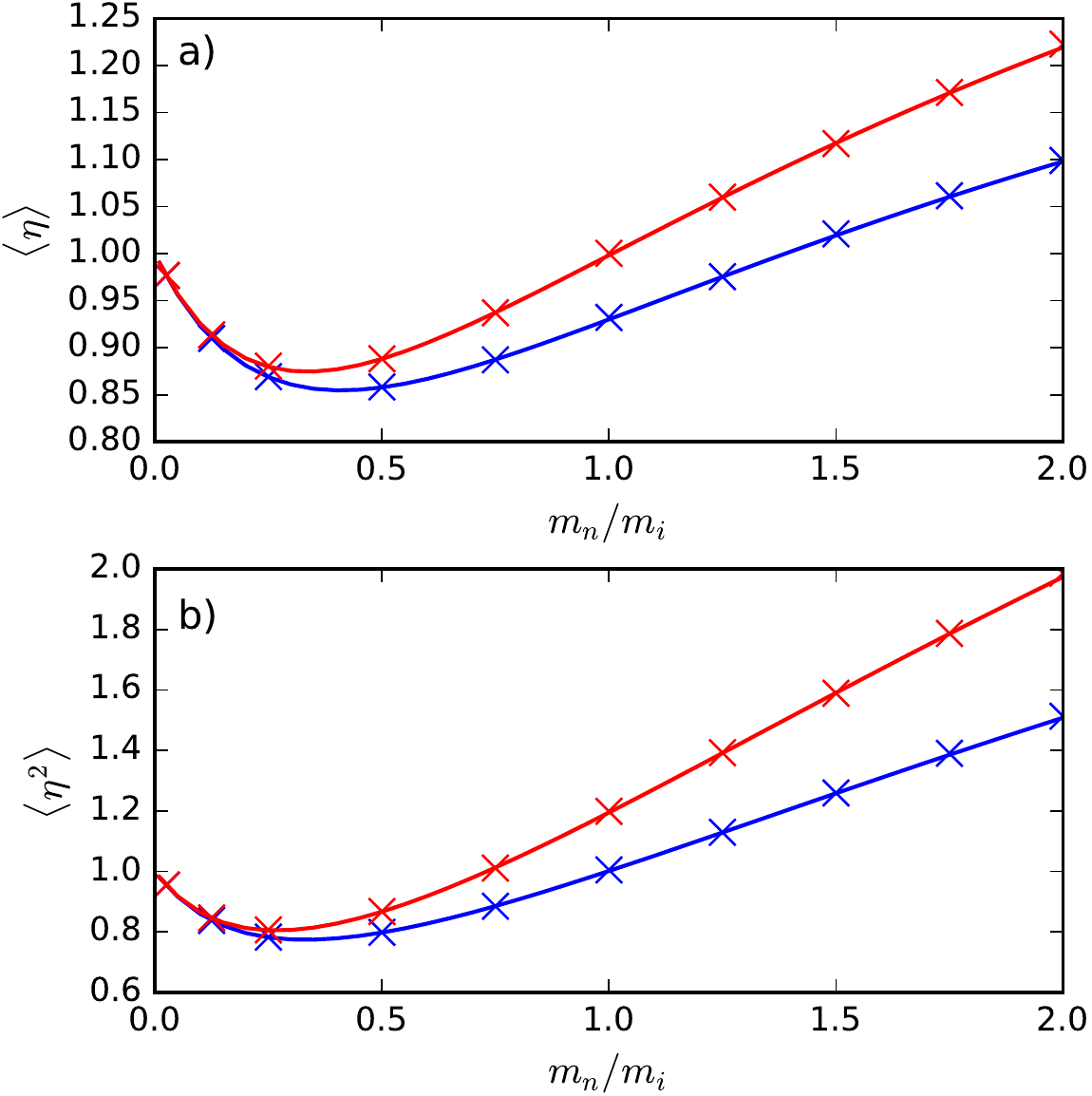}
    \caption{(a) A comparison of the mean value of $\eta$ extracted from numerical simulation (points) to the values calculated from the procedure outlined in the text (lines) for $q_R = 0.1$ (blue) and $q_R = 0.5$ (red). Points represent the result of 100'000 simulations, error bars corresponding to the standard error are not visible on this scale. (b) As (a), except for the mean value of $\eta^2$. In both cases, the expressions are calculated assuming that equipartition of energy applies between all three spatial degrees of freedom.
\label{fig:etaMeanVals}}
\end{figure}

We may make use of these distributions and the expression for $\eta$ to calculate $\langle\eta\rangle$ and $\langle\eta^2\rangle$ by averaging it in turn over each of these distributions.  The integrations over $\phi_{x,y,z}, x_1,x_2,x_3,\phi_\rho,\theta_\rho$  may be performed analytically, leaving only the integration over $\tau$ to be performed numerically. We find that the remaining function of $\tau$ is periodic and so integrating over a single period is sufficient to calculate $\langle \eta \rangle$ and $\langle \eta^2 \rangle$ in terms of the Mathieu parameters and the mass ratio. As shown in Fig.~\ref{fig:etaMeanVals}, the mean value calculated using this procedure and the stated distributions for the random variables involved is in excellent agreement with the values found from simulations. 

%%%%% Derivation of the heating coefficient

\subsection{Derivation of the heating coefficient $\kappa$} \label{section:zeroEnergy}

\begin{figure}
\centering
\includegraphics[width=0.9\columnwidth]{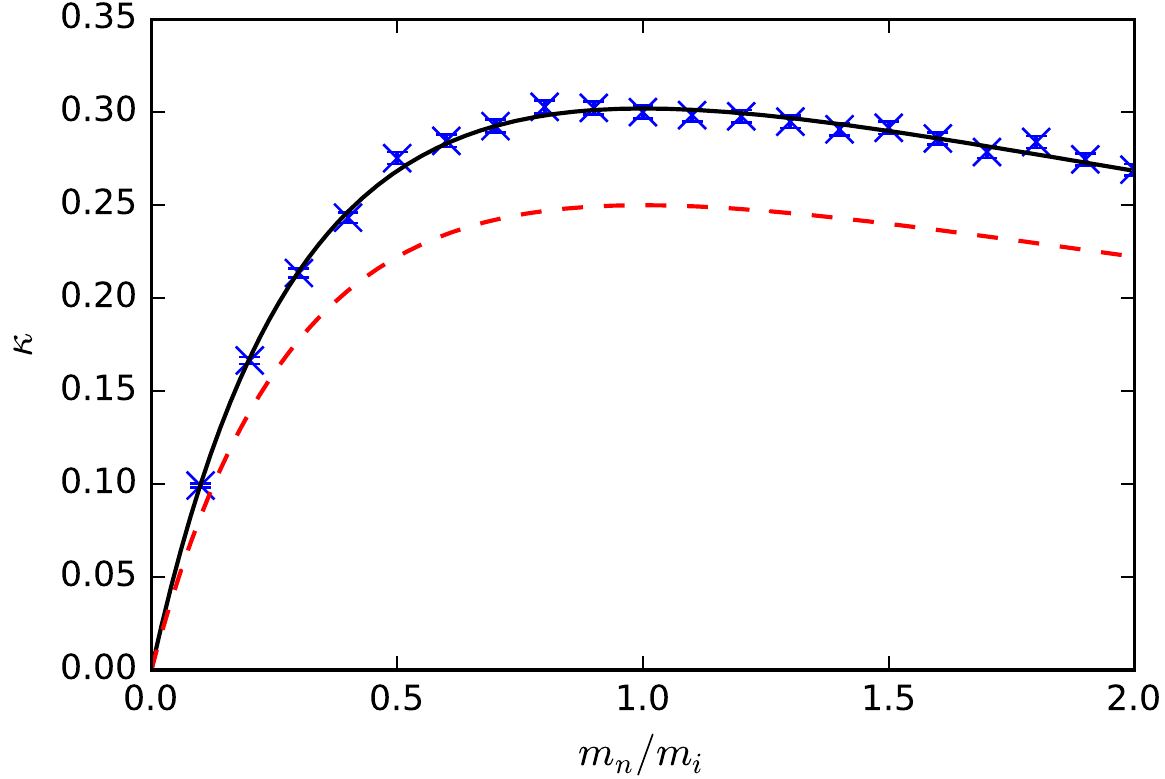}
\caption{The coefficient of thermal transfer $\kappa$ as a function of the mass ratio for $q=0.5$. The points give the value calculated from the mean energy from 10 000 numerical simulations of collisions of an ion initially at rest, with the error bars corresponding to the standard error on this value. The red dashed line is the theoretical approximation for the limit $q \rightarrow 0$, and the black solid line is the exact expression according to Eq. \eqref{eq:kappa}.}
\label{fig:kappaValue}
\end{figure}

This approach may also be used to calculate the rate of transfer between the thermal energy of the atoms and the ion, Eq.~(20) in the main text. By assuming an ion initially at the centre of the trap at rest, we may calculate the energy gained in a collision and apply equipartition to translate this to an increase in the temperature of the ion. Briefly, we take the Eq. \eqref{eq:appendixEnergyRelation} and set $E' = 3 k_b T', E = 0, \epsilon = \frac{3}{2} k_b T_a $ to obtain $T' = \kappa T_a$, where $\kappa$ is a function of the random variables $x_1,x_2,x_3,\phi_n,\theta_n,\tau$ as defined earlier, the mass ratio $\tilde{m}$ and the Mathieu parameters $a_j,q_j$. As outlined in the main text, we assume that $\kappa$ can be set to its mean value and so we average over the three uniform variables $x_j$ and the angular distributions of $\phi_n, \theta_n$. The result is,
\begin{align}
\kappa &= \frac{\tilde{m}}{3 (1 + \tilde{m})^2} \Biggr(  \sum_{j=x,y,x} \frac{c_{0,j}^2 \beta_j^2}{w_j^2} \big(\matc(a_j,q_j,\tau)^2 \notag \\
&+ \mats(a_j,q_j,\tau)^2  \big)   \Biggr) 
\end{align}
where $w_j^2$ is the Wronskian defined by $(\dot{\matc}(a_j,q_j,\tau) \mats(a_j,q_j,\tau)- \matc(a_j,q_j,\tau) \dot{\mats}(a_j,q_j,\tau))^2$, which is time-independent \cite{chen14a}. Furthermore, when averaged over a complete period, $\langle \matc(a_j,q_j,\tau)^2 + \mats(a_j,q_j,\tau)^2 \rangle = 1$, and so we obtain,
\begin{equation}\label{eq:kappa}
\kappa = \frac{\tilde{m}}{3 (1 + \tilde{m})^2} \left(  \sum_{j=x,y,x} \frac{c_{0,j}^2 \beta_j^2}{w_j^2}   \right) \approx \frac{\tilde{m}}{ (1 + \tilde{m})^2}
\end{equation}
where the approximation applies in the limit $q \rightarrow 0$, i.e. a time-independent trap.  In Fig.~\ref{fig:kappaValue}, we plot the calculated value of $\kappa$ for a time-dependent trap and the value extracted from numerical simulations for $q=0.5$ over a range of mass ratios, finding an excellent agreement between the two (black solid line). The approximate function for a time-independent trap (red dashed line) does not adequately describe $\kappa$ at this high value of $q$, but captures the dependence on the mass ratio.

%%%%% Numerical methods

\subsection{Numerical methods} \label{section:numericalMethods}

In order to generate numerical energy distributions for comparison with our statistical-mechanical models, we employ the Monte-Carlo simulation approach described by DeVoe \cite{devoe09a}. The ion's initial state is calculated by first generating a value for the secular amplitude of motion, $\tilde{A}_j$, for each axis from the thermal distribution,
\begin{equation} \label{eq:thermalAmplitudeDist}
\dist{\tilde{A}_j} = \tilde{A}_j m_i \beta_0 \omega_j^2 \exp \left[  -\frac{1}{2} \tilde{A}_j^2 m_i \beta_0 \omega_j^2  \right] ,
\end{equation}
 where $\omega_j = \beta_j \Omega/2$, and $\beta_0$ is the initial inverse temperature.  The inverse transform method may be applied to generate random values from this distribution by taking a uniform random number $u \in [0,1)$  and solving \cite{abramowitz70a},
\begin{equation}
\tilde{A_j} = \frac{1}{\omega_j}   \sqrt{\frac{2 k_B T}{m_i}   \ln \left(\frac{1}{1-u}\right)}.
\end{equation}
The secular phase $\phi_j$ is taken from a uniform distribution $[0,2\pi)$ and used with the Mathieu amplitude $A_j = \tilde{A}_j / c_{0,j}$ to calculate the initial position and velocity of the ion from Eqs.~\eqref{eq:ionMotion2Mathieu} and \eqref{eq:ionVelocity2Mathieu}. In this way, we ensure that the initial states are generated according to a fixed initial temperature and that the initial secular velocity is correctly correlated to the initial position. Since collisions occur at an energy-independent rate, they are described by Poisson statistics, and so the time between collisions may be sampled from an exponential distribution. The simulation is advanced directly from one collision to the next through use of the matrix propagator method to avoid errors introduced by numerical integration \cite{devoe09a}. 
Collisions are simulated by updating the ion's velocity components according to Eq.~\eqref{eq:collisionAtomVel2}. Velocities of atoms are drawn from a normal distribution at a fixed temperature, or set to zero to investigate only the multiplicative effects. Three uniformly distributed random numbers $\in [0,1)$ are required for the random rotation matrix, which is calculated as described in Ref.~\cite{arvo91a}. The simulation is then advanced to the next collision and the process repeats until a certain number of collisions have been simulated at which point the final secular energy is calculated.

The first two moments of the unknown distribution $\dist{\ln \eta}$, i.e. $\mu$ and $\sigma^2$,  are required as parameters in Eqs.~(16) and (18) of the main text. We may estimate these parameters from the empirical distribution for $\eta$, Eq.~(5) in the main text, or we may extract them from numerical simulations. The estimated value of $\sigma^2$ is typically only accurate to within $\pm 20\%$ and so these are instead extracted from simulations of single collisions with buffer-gas atoms at 0~K, for which $\eta = E'/E$. Performing a large number of such simulations allows the estimation of $\mu$ and $\sigma^2$ from the resulting distribution of $\eta$. These estimates do not take into account the deviation from equipartition, since they are calculated from a thermal state, but are sufficiently accurate to allow comparison of the derived distribution to the numerical results.

For comparisons to values estimated with Eq.~(24) and (27) in the main text, $\langle \beta \rangle$ and $n_T$ may also be found by maximum likelihood estimation (MLE) of numerical data. This has been shown to be more accurate than curve-fitting to binned data for Tsallis functions, avoids the issues of choosing an appropriate bin size, and requires only numerical root finding for one variable rather than a 2D optimization process \cite{shalizi07a}. The procedure for standard Tsallis functions with $k=0$ is described in Ref.~\cite{shalizi07a} and may be applied straightforwardly for other values of $k$.

Unless stated otherwise, numerical simulations are performed assuming an initial ion temperature $T_0$ of 1~mK, an arbitrary collision rate of 1000 s$^{-1}$, an RF frequency of $8 \times  2\pi$ MHz in a linear Paul trap with $q_x = 0.2$, $q_y = -0.2$, $q_z = 0.0$. $a_z$ is set such that the axial frequency is 100 kHz, and $a_{x,y}$ are equal to $-\frac{1}{2}a_z$. All computations were performed in Mathematica 10.2 using the internal implementation of the Mathieu functions and calculation of $\beta_j$. When necessary, coefficients for the Fourier series representation of the Mathieu functions were calculated using Miller's algorithm \cite{fruchting69a}. Figures presented are log-binned and represent the results of at least 10000 trials.

%%%% Laplace Method

\subsection{Approximate calculation of the ion energy distribution for a zero-temperature buffer gas} \label{sm:LaplaceMethod}

To evaluate the ion energy distribution for a zero-temperature buffer gas (Eq.~(17) in the main text), 
\begin{align} \label{eq:energyDistribution}
\dist{E_n} &=  \int_{\beta_n=0}^{\beta_n=\infty}  \frac{E_n^k \beta_n^{k+1} }{\Gamma(k)} e^{-E_n \beta_n} \notag \\
& \times \frac{1}{\sqrt{2 \pi n}  \sigma \beta_n} \exp[   -\frac{ (\ln \beta_n - \ln \beta_0 + n\mu)^2  }{2 n \sigma^2}  ]   \mathrm{d}\beta_n.
\end{align}
we use the Laplace method to find an approximate solution for $k=2$ following Ref.~\cite{asmussen16a}. Briefly, the integrand of Eq.~\eqref{eq:energyDistribution} has a maximum at the point $\beta_n = \hat{\beta}$,
\begin{equation}
\hat{\beta} = \beta_0 \exp \left(-n \mu + 2 n \sigma^2 - \mathcal{W}\left[\beta_0 E_n n \sigma^2 e^{2 n \sigma^2-\mu n}\right]  \right),
\end{equation}
where $\mathcal{W}$ is the Lambert-$\mathcal{W}$ function \cite{asmussen16a}. We define $g(E_n,\beta_n)$ to be the logarithm of the integrand of Eq.~\eqref{eq:energyDistribution} such that,
\begin{equation}
\dist{E_n} = \int_{\beta_n=0}^{\beta_n=\infty} \exp(g(E_n,\beta_n)) d \beta_n,
\end{equation}
and then replace $g(E_n,\beta_n)$ with its Taylor series to second order around the point $\beta = \hat{\beta}$. This leads to a Gaussian integral which can be analytically evaluated,
\begin{align} \label{eq:eDistNApprox}
&\dist{E_n} = \frac{\hat{\beta}^3 E_n^2 }{4 \sqrt{\hat{\beta} E_n n \sigma^2+1}}  \exp \left( -\hat{\beta} E_n \right)  \notag \\
& \times \left(\text{erf}  \left(\sqrt{\frac{\hat{\beta}E_n n \sigma^2+1}{2 n \sigma^2}}\right)  +1\right) \exp \left(  - \frac{n \sigma^2}{2} \left( \hat{\beta}E_n -2\right)^2  \right) ,
\end{align}
which is asymptotically correct for $E_n \rightarrow \infty$, since as $E_n$ increases, the integral becomes more sharply peaked around $\hat{\beta}$ and the approximation becomes more precise  \cite{asmussen16a}. The same method can be applied for an arbitrary value of $k$. 

\subsection{Derivation of the $\beta$ distribution for an ion in a buffer gas at finite temperature}
Briefly, following Ref.~\cite{sornette97a}, we take Eq.~(20), 
\begin{equation}
T_i = \eta_i T_{i-1} + \kappa T_a,
\end{equation}
and rewrite this as,
\begin{equation}
\frac{T_i-T_{i-1}}{T_{i-1}}=\eta_i-1+\kappa\frac{T_a}{T_{i-1}}.
\end{equation}
Using $(T_i-T_{i-1})/T_{i-1}\approx d\ln T / dt$, this expression can be converted into an overdamped Langevin equation for $x = \ln T$,
\begin{equation}
\frac{d x}{d t} = \mu + \hat{\eta}(t) + \kappa T_a e^{-x}  ,
\end{equation}
where $\mu=\langle\eta\rangle-1$ and $\eta$ has been decomposed into its mean $\langle \eta\rangle$ and a  fluctuating part $\hat{\eta}(t)$. This Langevin equation can be approximated by a Fokker-Planck equation which in steady state is given by,
\begin{equation} \label{eq:thermalFokkerPlanck}
\frac{\sigma^2}{2} \frac{d^2}{d x^2} f_x(x) - \frac{d}{dx}\left[   (     \mu + \kappa T_a e^{-x}    ) f_x(x) \right] = 0 .
\end{equation}
The boundary conditions are fixed by $f_T(0)\rightarrow0$ and $f_T(\infty)\rightarrow0$, corresponding to   $f_x(x) \rightarrow 0$ for $x\rightarrow \pm \infty$. The corresponding solution of Eq. (\ref{eq:thermalFokkerPlanck}) is then given by,
\begin{equation}
\dist{x} = A \exp \left(-\frac{2}{\sigma^2}   (\kappa T_0 e^{-x}-\mu x)\right) ,
\end{equation}
where $A$ is a normalization constant. Proceeding directly to $\beta = e^{-x}$ and normalizing for $\mu < 0$ we find, 
\begin{equation}
\dist{\beta} = \frac{1}{\beta \Gamma (\nu)}   e^{-\frac{\beta \nu}{b}} \left(\frac{\beta \nu}{b}\right)^{\nu} 
\end{equation}
where $\nu =  -\frac{2\mu}{\sigma^2} $ and $b = \frac{ - \mu  }{ k_B T_a \kappa} $. This is a gamma distribution, in agreement with the result obtained in Ref.~\cite{biro05a} for multiplicative fluctuations with an additive constant. Due to the approximations used in the derivation, the forms given for the two parameters $\nu$ and $b$ correspond to those expected for a log-normal distribution of $\eta$, which is not the case for the present system. We therefore replace $\nu$ and $b$ with $n_T$ and $\langle \beta \rangle$ and calculate values for these appropriate for the observed form of $\dist{\eta}$ as described in the main text.

\end{document}